\documentclass[aps,showkeys,twocolumn,preprintnumbers,showpacs,amsmath,amssymb,superscriptaddress,floatfix,nofootinbib]{revtex4}
\usepackage{graphicx,color,dcolumn,booktabs,bm}
\usepackage{longtable,lscape}
\usepackage{txfonts}
\usepackage{overpic}
\usepackage{epsfig}
\usepackage{amssymb}
\usepackage{rotating}
\usepackage{epstopdf}
\usepackage{indentfirst}
\usepackage{feynmf}   %{feynmp}
\usepackage{slashed}  %for Feynman symbols
\usepackage{cases}
\usepackage{color}
\usepackage{multirow}
\usepackage{graphicx,color,dcolumn,booktabs,bm}
\usepackage{cases}
\usepackage{array}

\graphicspath{{Figures/}} %
\usepackage[colorlinks, citecolor=blue,anchorcolor=red,menucolor=red, linkcolor=red,filecolor=red,runcolor=red,urlcolor=blue,frenchlinks=red]{hyperref}

\begin{document}
\title{Strong decays of the low-lying bottom strange baryons }
\author{Hui-Zhen He}
\affiliation{  Department
of Physics, Hunan Normal University,  Changsha 410081, China }

\affiliation{ Synergetic Innovation
Center for Quantum Effects and Applications (SICQEA), Changsha 410081,China}

\affiliation{  Key Laboratory of
Low-Dimensional Quantum Structures and Quantum Control of Ministry
of Education, Changsha 410081, China}

\author{Wei Liang}
\affiliation{  Department
of Physics, Hunan Normal University,  Changsha 410081, China }

\affiliation{ Synergetic Innovation
Center for Quantum Effects and Applications (SICQEA), Changsha 410081,China}

\affiliation{  Key Laboratory of
Low-Dimensional Quantum Structures and Quantum Control of Ministry
of Education, Changsha 410081, China}

\author{Qi-Fang L\"u \footnote{Corresponding author} } \email{lvqifang@hunnu.edu.cn} %
\affiliation{  Department
of Physics, Hunan Normal University,  Changsha 410081, China }

\affiliation{ Synergetic Innovation
Center for Quantum Effects and Applications (SICQEA), Changsha 410081,China}

\affiliation{  Key Laboratory of
Low-Dimensional Quantum Structures and Quantum Control of Ministry
of Education, Changsha 410081, China}

\author{Yu-Bing Dong}
\affiliation{Institute of High Energy Physics, Chinese Academy of Sciences, Beijing 100049, China}
\affiliation{Theoretical Physics Center for Science Facilities (TPCSF), Chinese Academy of Sciences, Beijing 100049, China}
\affiliation{School of Physical Sciences, University of Chinese Academy of Sciences, Beijing 101408, China}

\begin{abstract}
In this work, the strong decay behaviors of the $\lambda-$mode low-lying $\Xi_{b}$ and $\Xi_{b}^\prime$ baryons are investigated within the $^3P_0$ model. Our results suggest that all of the low-lying $\bar 3_F$ states $\Xi_b(2S)$, $\Xi_b(1P)$, and $\Xi_b(1D)$ have small decay widths of less than 20 MeV, and these states have good potentials to be observed in the $\Xi_b^\prime \pi$ and $\Xi_b^{\prime *} \pi$ invariant masses. Further, most of the $6_F$ multiplets are relatively narrow and may decay into the $\Xi_b \pi$, $\Xi_b^\prime \pi$, $\Xi_b^{\prime *} \pi$, and $\Lambda_b \bar K$ final states. Considering the masses and strong decay behaviors, we can assign the newly observed $\Xi_{b}(6100)$ resonance as the $\Xi_b(1P)$ state with $J^P=3/2^-$ and interpret the $\Xi_{b}(6227)$ structure as the $J^P=3/2^-$ $\Xi_b^\prime(1P)$ state with a proper mixing angle. We expect that our predictions for these excited bottom strange baryons will provide helpful information for future experimental research.

\end{abstract}

\keywords{bottom strange baryons, strong decays, $^3P_0$ model}
\pacs{14.20.Mr, 13.30.Eg, 12.39.-x}

\maketitle

\section{Introduction}

Investigating the properties of baryons and searching for missing states is a vast and significant topic in hadronic physics. Among the baryons, the singly bottom baryons are particularly interesting, since the heavy quark symmetry is preserved well in them, and their mass splitting is significantly smaller than that in the case of light and charmed baryons. In the past years, many singly bottom baryons were experimentally observed, thereby providing us good opportunities to study their mass spectra and inner structures. In 2018, the Large Hadron Collider beauty (LHCb) Collaboration observed the $\Sigma_{b}(6097)^{\pm}$ structures in the $\Lambda_{b}^{0} \pi^{\pm}$ invariant masses~\cite{Aaij:2018tnn} and announced the observation of a bottom strange baryon $\Xi_b(6227)$ in both $\Lambda_b \bar K$ and $\Xi_b \pi$ decay modes~\cite{Aaij:2018yqz}. In 2019, the Collaboration discovered two resonances $\Lambda_{b}(6146)$ and $\Lambda_{b}(6152)$ in the $\Lambda_{b}^{0} \pi^{+} \pi^{-}$ decay mode~\cite{Aaij:2019amv}. In 2020, the Collaboration observed four narrow peaks $\Omega_{b}(6316)^{-}$, $\Omega_{b}(6330)^{-}$, $\Omega_{b}(6340)^{-}$ and $\Omega_{b}(6350)^{-}$ in the $\Xi_{b}^{0} K^{-}$ final state~\cite{Aaij:2020cex}. In addition, in 2020, the Compact Muon Solenoid (CMS) Collaboration found a broad structure in the $\Lambda_{b}^{0} \pi^{+} \pi^{-}$ final state within the energy region of $6040 \sim 6100$ MeV~\cite{Sirunyan:2020gtz}, which was subsequently confirmed by the LHCb Collaboration~\cite{Aaij:2020rkw}. Moreover, a new bottom–strange baryon $\Xi_{b}(6100)^-$ was found very recently by the CMS Collaboration in the $\Xi_b^- \pi^+ \pi^-$ invariant mass spectrum ~\cite{Sirunyan:2021vxz}.

The dramatic progress in the above experiments has attracted much attention in the field of theoretical physics and has helped establish the low-lying spectra for the singly bottom baryons more efficiently~\cite{Wang:2019uaj,Chen:2019ywy,Cui:2019dzj,Chen:2018vuc,Wang:2018fjm,Yang:2018lzg,Aliev:2018vye,Jia:2019bkr,Yang:2019cvw,Liang:2019aag,Liang:2020hbo,Liang:2020kvn,Xiao:2020oif,Wang:2020pri,Karliner:2020fqe,Xu:2020ofp,Wang:2020mxk,Arifi:2020yfp,Ortiz-Pacheco:2020hmj,Bijker:2020tns,Xiao:2020gjo,Chen:2018orb,Chen:2020mpy}. For instance, in the $\Lambda_b$ sector, the $\Lambda_b(5912)$ and $\Lambda_b(5920)$ have been assigned as the $\Lambda_b(1P)$ doublet; two structures $\Lambda_b(6146)$ and $\Lambda_b(6152)$ can be clarified as belonging to the $\Lambda_b(1D)$ doublet, and the $\Lambda_b(6072)$ may be the $\Lambda_b(2S)$ state or the lowest $\rho-$mode state~\cite{Wang:2017kfr,Chen:2019ywy,Liang:2019aag,Wang:2019uaj,Ebert:2011kk,Roberts:2007ni,Chen:2014nyo,Ebert:2007nw,Capstick:1986bm,Aaij:2020rkw,Liang:2020kvn}. Hence, the low-lying $\Lambda_b$ spectrum is almost established. Further, with regard to the $\Omega_b$ system, most of the $P-$wave states are believed to have been found~\cite{Liang:2020hbo,Chen:2020mpy,Xiao:2020oif,Wang:2020pri,Karliner:2020fqe,Ortiz-Pacheco:2020hmj,Xu:2020ofp}. However, unlike the $\Lambda_b$ and $\Omega_b$ systems, the spectra of bottom strange baryons are far from established.

A bottom strange baryon is composed of a heavy bottom quark, a light up or down quark, and a strange quark. In the constituent quark model, the bottom-strange baryons can be divided into two families, namely, the antisymmetric flavor configuration $\bar 3_F$ and symmetric flavor configuration $6_F$. Theoretically, the $\bar 3_F$ states are denoted as $\Xi_b$ baryons, and the $6_F$ multiplets, as $\Xi_b^\prime$ states. Unlike the $\Lambda_b$ and $\Sigma_b$ states with different isospins, the $\Xi_b$ and $\Xi_b^\prime$ baryons may have the same quantum numbers and cannot be distinguished experimentally. Hence, investigating and determining their inner structures both experimentally and theoretically is a complicated and challenging task.

Until now, only two excited structures $\Xi_b(6227)$ and $\Xi_b(6100)$ for the bottom strange baryons have been observed experimentally. $\Xi_b(6227)$ was investigated in several theoretical works. It is usually regarded as a $\Xi_b^\prime(1P)$ state with $J^P=3/2^-$ or $J^P=5/2^-$~\cite{Chen:2018orb,Wang:2018fjm,Jia:2019bkr,Cui:2019dzj,Azizi:2020azq}. Since it lies below the $\Sigma_b \bar K$ threshold, molecular assignment is also possible~\cite{Lu:2014ina,Huang:2018bed,Yu:2018yxl,Nieves:2019jhp,Zhu:2020lza,Wang:2020vwl}. For the $\Xi_b(6100)$, which was observed recently, the CMS Collaboration suggests that it should be a $\Xi_b(1P)$ state with $J^P=3/2^-$, but its inner structure has not been theoretically examined so far. Thus, it is necessary to study the strong decay behaviors of $\Xi_b(6100)$ before a firm conclusion can be achieved.

Although the constituent quark models have predicted many bottom–strange baryons for a long time, theoretical studies on their strong decay behaviors are relatively few~\cite{Wang:2018fjm,Yao:2018jmc,Wang:2017kfr,Chen:2018orb,Azizi:2020azq,Cui:2019dzj}. Since the predictions depend on the choice of models and parameters, it is crucial to employ a unified framework to describe the whole singly heavy baryons. In this work, we adopted the widely accepted $^3P_0$ model to systematically calculate the strong decays for the $\lambda-$mode low-lying bottom–strange baryons. Our present parameters have been extensively and successfully used for the $\Lambda_c$, $\Sigma_c$, $\Xi_c$, $\Xi_c^{\prime}$, $\Lambda_b$, $\Sigma_b$, and $\Omega_b$ systems~\cite{Lu:2018utx,Liang:2019aag,Lu:2019rtg,Liang:2020hbo,Lu:2020ivo,Liang:2020kvn}. Hence, we expect that these parameters should be suitable for the excited $\Xi_b$ and $\Xi_b^\prime$ states. Our results suggest that the newly observed $\Xi_{b}(6100)$ resonance can be regarded as the $\Xi_b(1P)$ state with $J^P=3/2^-$, and the $\Xi_{b}(6227)$ structure can be interpreted as the $J^P=3/2^-$ $\Xi_b^\prime(1P)$ state. Owing to the relatively narrow widths, there is good potential for observing most of the low-lying $\Xi_b$ and $\Xi_b^\prime$ states in future experiments.

The rest of this paper is organized as follows. Sec.~\ref{model} provides a brief introduction of the $^3 P_0$ model and notations. The strong decay behaviors of the low-lying $\Xi_b$ and $\Xi_{b}^\prime$ states are calculated and discussed in Sec.~\ref{low-lying}. A short summary is given in the last section.

\section{$^3P_0$ Model and notations}{\label{model}}

In this work, we adopted the $^3P_0$ model to calculate the two-body strong decays for the low-lying $\Xi_b$ and $\Xi_b^\prime$ states. In the $^3P_0$ model, a quark–antiquark pair with a quantum number $J^{PC}$ =$0^{++}$ is created from vacuum, and falls apart into the final states.~\cite{micu,Colglazier:1970vx}. This model has been widely employed to study the OZI-allowed strong decays of various hadronic systems, and considerable advances were achieved~\cite{micu,Colglazier:1970vx,3p0model1,3p0model2,3p0model4,3p0model5,3p0model6,Chen:2007xf,Zhao:2016qmh,Ye:2017dra,Chen:2017gnu,Chen:2016iyi,Lu:2014zua,Lu:2016bbk,Ferretti:2014xqa,Godfrey:2015dva,Segovia:2012cd,Mu:2014iaa,Lu:2018utx,Guo:2019ytq,Lu:2019rtg,Lu:2020ivo,Ye:2017yvl}. In this section, we briefly introduce this model. In the nonrelativistic limit, to describe a decay process $A\rightarrow BC$, the transition operator $T$ in the $^3P_0$ model can be written as
\begin{eqnarray}
T&=&-3\gamma\sum_m\langle 1m1-m|00\rangle\int
d^3\boldsymbol{p}_4d^3\boldsymbol{p}_5\delta^3(\boldsymbol{p}_4+\boldsymbol{p}_5)\nonumber\\&&\times {\cal{Y}}^m_1\left(\frac{\boldsymbol{p}_4-\boldsymbol{p}_5}{2}\right)\chi^{45}_{1,-m}\phi^{45}_0\omega^{45}_0b^\dagger_{4i}(\boldsymbol{p}_4)d^\dagger_{4j}(\boldsymbol{p}_5),
\end{eqnarray}
where $\gamma$ is a dimensionless $q_4\bar{q}_5$ pair production strength, and $\boldsymbol{p}_4$ and $\boldsymbol{p}_5$ are the momenta of the created quark $q_4$ and antiquark $\bar{q}_5$, respectively. Here, $i$ and $j$ are the color indices of the created quark and antiquark, respectively. $\phi^{45}_{0}=(u\bar u + d\bar d +s\bar s)/\sqrt{3}$, $\omega^{45}=\delta_{ij}$, and $\chi_{{1,-m}}^{45}$ are the flavor singlet, color singlet, and spin triplet wave functions of $q_4\bar{q}_5$, respectively. The ${\cal{Y}}^m_1(\boldsymbol{p})\equiv|p|Y^m_1(\theta_p, \phi_p)$ is the solid harmonic polynomial that reflects the $P-$wave momentum space distribution of the created quark pair.

\begin{figure}[!htbp]
\includegraphics[scale=0.5]{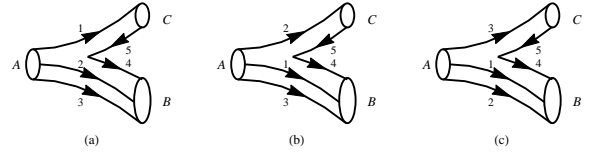}
\vspace{0.0cm} \caption{Decay process $A\to B+C$ for a bottom strange baryon in the $^3P_0$ model.}
\label{qpc}
\end{figure}

For the strong decay of a bottom strange baryon, there exist three possible rearrangements:
\begin{eqnarray}
A(q_1,s_2,b_3)+P(q_4,\bar q_5)\to B(s_2,q_4,b_3)+C(q_1,\bar q_5),\\
A(q_1,s_2,b_3)+P(q_4,\bar q_5)\to B(q_1,q_4,b_3)+C(s_2,\bar q_5),\\
A(q_1,s_2,b_3)+P(q_4,\bar q_5)\to B(q_1,s_2,q_4)+C(b_3,\bar q_5).
\end{eqnarray}
These three ways of recouplings are also presented in Figure~\ref{qpc}. It can be seen that the
first and second ones indicate the heavy baryon-light meson channels, and the last one denotes the light
baryon-heavy meson decay mode.

With the transition operator $T$, the helicity amplitude ${\cal{M}}^{M_{J_A}M_{J_B}M_{J_C}}$ is defined as
\begin{eqnarray}
\langle
BC|T|A\rangle=\delta^3(\boldsymbol{P}_A-\boldsymbol{P}_B-\boldsymbol{P}_C){\cal{M}}^{M_{J_A}M_{J_B}M_{J_C}}.
\end{eqnarray}
The explicit formula for the helicity amplitude can be found in Refs.~\cite{Chen:2016iyi,Chen:2007xf,Ye:2017yvl,Lu:2018utx,Liang:2019aag}. Then, the decay width of $A\to BC$ process can be obtained easily as
\begin{eqnarray}
\Gamma= \pi^2\frac{p}{M^2_A}\frac{1}{2J_A+1}\sum_{M_{J_A},M_{J_B},M_{J_C}}|{\cal{M}}^{M_{J_A}M_{J_B}M_{J_C}}|^2,
\end{eqnarray}
where $p=|\boldsymbol{p}|$ is the momentum of the final hadrons in the center of mass system.

The notations of the relevant initial states and the masses predicted by the relativistic quark model~\cite{Ebert:2011kk} are listed in Table~\ref{bmass}. First, most of the theoretical works only focus on the $\lambda-$mode excitation in the study of mass spectra~\cite{Ebert:2011kk,Roberts:2007ni,Chen:2014nyo}. Relatively fewer studies have explored $\rho-$mode spectra. Next, a series of singly heavy baryons were observed experimentally, and they are almost assigned as the $\lambda-$mode excitations owing to their masses and strong decays~\cite{Liang:2019aag,Liang:2020kvn,Wang:2019uaj,Chen:2019ywy,Chen:2018vuc,Chen:2018orb,Wang:2018fjm,Yang:2019cvw}. Meanwhile, no $\rho-$mode excitation has been confirmed either theoretically or experimentally until now. Hence, only the $\lambda-$mode low-lying states are considered here, and the $\rho-$mode quantum numbers $n_{\rho}=l_{\rho}=0$ are neglected. The $\Xi_{b}(6100)$ and $\Xi_b(6227)$ resonances are tentatively assigned as the $\Xi_b(1P)$ and $\Xi_b^\prime(1P)$ states, respectively, and their strong decay behaviors are discussed. For the final ground states, the masses of these states are taken from the Review of Particle Physics~\cite{pdg}. All the parameters in the $^3P_0$ model used here are the same as those used in our previous works~\cite{Lu:2018utx,Liang:2019aag,Lu:2019rtg,Liang:2020hbo,Lu:2020ivo,Liang:2020kvn}. These parameters have been well tested, and they describe the strong decay behaviors of various singly heavy baryons successfully. More discussions on these parameters can be found in Refs.~\cite{Liang:2019aag,Liang:2020hbo}.

\begin{table}[htb]
\begin{center}
\caption{ \label{bmass} Notations, quantum numbers, and masses of the initial baryons. $n_\lambda$ and $l_\lambda$ represent the nodal quantum number and orbital angular momentum, respectively, between the heavy quark and light quark system. $S_\rho$ stands for the total spin of the two light quarks; $L$ is the total orbital angular momentum; $j$ represents the total angular momentum of $L$ and $S_\rho$; $J$ is the total angular momentum; $P$ is the parity. The masses are taken from the theoretical predictions of the relativistic quark model~\cite{Ebert:2011kk}. The values are in MeV. }
\renewcommand\arraystretch{1.8}

\begin{tabular*}{8.6cm}{@{\extracolsep{\fill}}*{8}{p{0.7cm}<{\centering}}}
\hline\hline
State &$n_{\lambda}$ & $l_{\lambda}$ & $L$	& $S_{\rho}$ &$j$	& $J^P$ & Mass\\
\hline
$\Xi_{b}(2S)$	&	1	&			0	&	0	&		0	&	0	&	$\frac{1}{2}^+$ & 6266 \\
$\Xi_{b}^{\prime}(2S)$	&	1	&			0	&	0	&		1	&	1	&	$\frac{1}{2}^+$ &	6329 \\
$\Xi_{b}^{\prime*}(2S)$	&	1	&			0	&	0	&		1	&	1	&	$\frac{3}{2}^+$ &	6342 \\
$\Xi_{b1}(\frac{1}{2}^-)$ &	0	& 		1	&	1	&		0	&	1	&	 $\frac{1}{2}^-$ &	6120 \\
$\Xi_{b1}(\frac{3}{2}^-)$ &	0	& 		1	&	1	&		0	&	1	&	$\frac{3}{2}^-$ &	6130 \\
$\Xi_{b0}^{\prime}(\frac{1}{2}^-)$	&	0		 &	1	&	1	&	1	&	0	&	$\frac{1}{2}^-$ &6233 \\
$\Xi_{b1}^{\prime}(\frac{1}{2}^-)$	&	0		 &	1	&	1	&	1	&	1	&	$\frac{1}{2}^-$ &6227 \\
$\Xi_{b1}^{\prime}(\frac{3}{2}^-) $ &	0	 	&	1	&	1	&	1	&	1	&	$\frac{3}{2}^-$ &	6234 \\
$\Xi_{b2}^{\prime}(\frac{3}{2}^-)$ &	0	& 		1	&	1	&	1	&	2	&	$\frac{3}{2}^-$ &	6224 \\
$\Xi_{b2}^{\prime}(\frac{5}{2}^-)$	&	0			&	1	&	1	&	1	&2	&	$\frac{5}{2}^-$ &	6226 \\
$\Xi_{b2}(\frac{3}{2}^+)$	&	0	&			2	&	2	&		0	&	2	&	$\frac{3}{2}^+$ & 6366 \\
$\Xi_{b2}(\frac{5}{2}^+)$	&	0	&			2	&	2	&		0	&	2	&	$\frac{5}{2}^+$ &	6373 \\
$\Xi_{b1}^{\prime}(\frac{1}{2}^+)$ &	0	& 		2	&	2	&	1	&	1	&	$\frac{1}{2}^+$ &	6447 \\
$\Xi_{b1}^{\prime}(\frac{3}{2}^+) $ &	0	 	&	2	&	2	&	1	&	1	&	 $\frac{3}{2}^+$ &	6459\\
$\Xi_{b2}^{\prime}(\frac{3}{2}^+)$ &	0	& 		2	&	2	&	1	&	2	&	 $\frac{3}{2}^+$ &	6431 \\
$\Xi_{b2}^{\prime}(\frac{5}{2}^+)$	&	0			&	2&	2&	1&	2	&	$\frac{5}{2}^+$ &	6432\\
$\Xi_{b3}^{\prime}(\frac{5}{2}^+)$	&	0	&			2	&	2	&	1	&	3	&	$\frac{5}{2}^+$ &	6420\\
$\Xi_{b3}^{\prime}(\frac{7}{2}^+)$	&	0	&			2	&	2	&	1&	3	&	$\frac{7}{2}^+$ &	6414\\
\hline\hline
\end{tabular*}
\end{center}
\end{table}

\section{Strong decays}{\label{low-lying}}
\subsection{$\Xi_b(2S)$ state}

In the traditional quark model, there exists a $\lambda$-mode $\Xi_{b}(2S)$ state with predicted mass of 6266 MeV.
With this mass, the strong decays of this state are calculated and listed in Table~\ref{B2s}. The total decay width is about 7 MeV, and the dominating decay modes are $\Xi_{b}^{\prime} \pi$ and $\Xi_{b}^{\prime *} \pi$. The branching ratios for these decay modes are predicted as 
\begin{equation}
Br(\Xi^{\prime}_{b} \pi, \Xi^{\prime *}_{b} \pi)=37.9\%,62.1\%,
\end{equation}
which are independent of the overall strength $\gamma$. Our predicted branching ratios are consistent with those of the potential model~\cite{Chen:2018orb}, while our calculated total decay width is narrower. These findings indicate that the $\Xi_b(2S)$ state is most likely to be observed in the $\Xi_{b}^{\prime} \pi$ and $\Xi_{b}^{\prime *} \pi$ final states in future experiments.

\begin{table}[!htbp]
\begin{center}
\caption{\label{B2s}  Theoretical predictions of the strong decays for the $\Xi_b(2S)$, $\Xi_b(1P)$ and $\Xi_b(1D)$ states
 in MeV.}
\renewcommand{\arraystretch}{1.5}
\footnotesize
\begin{tabular*}{8.6cm}{@{\extracolsep{\fill}}*{6}{p{1.3cm}<{\centering}}}
\hline\hline
Mode &$\Xi_b(2S)$&	$\Xi_{b1}(\frac{1}{2}^-)$	&	$\Xi_{b1}(\frac{3}{2}^-)$&$\Xi_{b2}(\frac{3}{2}^+)$	&	$\Xi_{b2}(\frac{5}{2}^+)$\\\hline
$\Xi_{b}^{\prime} \pi$&2.65&17.59&0.01 &2.89&0.06\\
$\Xi_{b}^{\prime *} \pi$&4.34&$3\times10^{-3}$&15.76&0.58&3.27\\
$\Sigma_{b} \bar K$&$-$&$-$&$-$&2.75&0.01\\
$\Sigma_{b}^{*} \bar K$&$-$&$-$&$-$&0.29&2.25\\
Total width&6.99&17.59&15.77&6.51&5.59\\
\hline\hline
\end{tabular*}
\end{center}
\end{table}

\subsection{$\Xi_b(1P)$ states}

In the constituent quark model, there are two $\Xi_b(1P)$ states, and their predicted masses are around 6100 MeV. The strong decays of these states are estimated using these masses, and the values are listed in Table~\ref{B2s}. Our results show that the total decay widths of these two $j=1$ states are similar. To distinguish them, the branching ratios should be discussed further. The dominant decay modes are $\Xi^{\prime}_{b} \pi$ and $\Xi_{b}^{\prime*} \pi$ for the $\Xi_{b1}(\frac{1}{2}^-)$ and $\Xi_{b1}(\frac{3}{2}^-)$ states, respectively. The two $\Xi_b(1P)$ states exhibit entirely different decay behaviors, which can help us distinguish them in future experiments.

\begin{figure}[htb]
\includegraphics[scale=0.5]{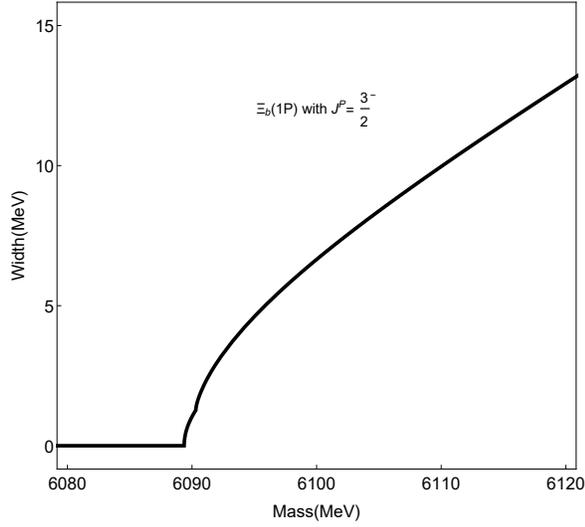}
\vspace{0.0cm} \caption{Strong decay behaviors for the $\Xi_{b1}(\frac{3}{2}^-)$ state versus its initial mass. The black line corresponds to the total decay width.}
\label{xib1p}
\end{figure}

Very recently, the CMS Collaboration announced the observation of the $\Xi_b(6100)^-$ resonance in the $\Xi_b^- \pi^+ \pi^-$ invariant mass spectrum and suggested that it should be a $\Xi_b(1P)$ state with spin and parity quantum numbers $J^P=3/2^-$~\cite{Sirunyan:2021vxz}. Assigning the $\Xi_b(6100)$ as the $\Xi_b(1P)$ states with $S_{\rho}$=0 tentatively, we adopted a mass of 6100.3 MeV to recalculate the strong decays. The estimated total width of the $J^P=3/2^-$ state is several MeV, which is slightly larger than the value from experimental observations. From the data in Table~\ref{B1P6100}, we can see that the dominant decay mode for the $\Xi_{b1}(\frac{3}{2})^{-}$ state is $\Xi_{b}^{\prime*} \pi$, and its decay width is almost same as the total width. Here, we emphasize that the $^3P_0$ model cannot give us precise results owing to the relatively small phase space, but the results represent the order of magnitude. Indeed, the total decay width is sensitive to the mass of the initial state, which is shown in Figure.~\ref{xib1p}. In agreement with our work, previous works too predicted a narrow decay width for the $J^P=3/2^-$ $\Xi_b(1P)$ states~\cite{Wang:2017kfr,Chen:2018orb}. Moreover, two $\Xi_c(1P)$ states, which are narrow resonances with total widths of several MeV, are listed in the Review of Particle Physics~\cite{pdg}. Then, the heavy quark symmetry suggests that the two $\Xi_b(1P)$ states with $S_{\rho}$=0 should be also narrow, which agrees with the total width of the $\Xi_b(6100)$ structure. With the observation in the $\Xi_{b}^{\prime*} \pi$ decay mode, the quantum numbers $J^P=3/2^-$ for the $\Xi_b(6100)$ resonance are favored.

\begin{table}[!htbp]
\begin{center}
\caption{\label{B1P6100} Strong decay widths of the $\Xi_b(6100)$ as $\Xi_b(1P)$ states in MeV.}
\renewcommand{\arraystretch}{1.5}
\begin{tabular*}{8.6cm}{@{\extracolsep{\fill}}*{3}{p{2.8cm}<{\centering}}}
\hline\hline
Mode	&	$\Xi_b(6100)$ as $\Xi_{b1}(\frac{1}{2}^-)$	&	$\Xi_b(6100)$ as $\Xi_{b1}(\frac{3}{2}^-)$	\\	\hline	
$\Xi_{b}^{\prime} \pi$&12.02&$1\times10^{-3}$\\
$\Xi_{b}^{\prime*} \pi$&$2\times10^{-4}$&6.76\\
Total width&$12.02$&$6.76$ \\
\hline\hline
\end{tabular*}
\end{center}
\end{table}

\subsection{$\Xi_b(1D)$ states}
In the constituent quark model, there exist two $\lambda-$mode $D-$wave excited states, which are denoted as $\Xi_ {b2}(\frac{3}{2}^+)$ and $\Xi_ {b2}(\frac{5}{2}^+)$. From Table~\ref{bmass}, the predicted masses of these two states are around 6370 MeV. The calculated values of the strong decay widths of these $D-$wave states are listed in Table~\ref{B2s}. The total decay width of the $J^P = 3/2^+$ state is 6.51 MeV, and the main decay modes are $\Xi_{b}^{\prime} \pi$ and $\Sigma_{b} \bar K$. The explicit branching ratios are as follows:
\begin{equation}
Br(\Xi^{\prime}_{b} \pi, \Xi_{b}^{\prime*} \pi, \Sigma_{b} \bar K, \Sigma_{b}^{*} \bar K) =44.4\%, 8.9\%, 42.2\%, 4.5\%.
\end{equation}
For the $J^P = 5/2^+$ state, the total decay width is predicted as 5.59 MeV, and the dominating decay modes are $\Xi_{b}^{\prime*} \pi$ and $\Sigma_{b}^{*} \bar K$. The branching ratios can be obtained as follows:
\begin{equation}
Br(\Xi^{\prime}_{b} \pi, \Xi_{b}^{\prime*} \pi, \Sigma_{b} \bar K, \Sigma_{b}^{*} \bar K) =1.1\%, 58.5\%, 0.1\%,40.3\%.
\end{equation}
Although the total decay widths for these two $D-$wave states are similar, the significantly different branching ratios can help us distinguish them. Besides, our calculations agree roughly with the predictions obtained using a chiral quark model~\cite{Yao:2018jmc}, and we hope future experiments can search for these $D-$wave doublets as well as their non-strange partners $\Lambda_b(6146)$ and $\Lambda_b(6152)$.

\subsection{$\Xi_b^\prime(2S)$ states}
From Table~\ref{bmass}, the predicted masses of $\Xi_b^{\prime}(2S)$ and $\Xi_b^{\prime*}(2S)$ states are 6329 and 6342 MeV, respectively, where the mass gap is significantly small. Their strong decays were estimated with these masses. The values are listed in Table~\ref{T2S}. The predicted widths for $\Xi_b^{\prime}(2S)$ and $\Xi_b^{\prime*}(2S)$ states are 33.89 and 35.64 MeV, respectively. The branching ratios are predicted as
\begin{eqnarray}
Br(\Xi_{b} \pi, \Xi^{\prime}_{b} \pi, \Xi^{\prime*}_{b} \pi, \Lambda_{b} \bar K, \Sigma_{b} \bar K, \Sigma_{b}^{*} \bar K) = \nonumber \\
32.6\%,19.1\%,8.3\%,34.6\%,5.3\%,0.1\%
\end{eqnarray}
for the $\Xi_b^{\prime}(2S)$ state, and
\begin{eqnarray}
Br(\Xi_{b} \pi, \Xi^{\prime}_{b} \pi, \Xi^{\prime*}_{b} \pi, \Lambda_{b} \bar K, \Sigma_{b} \bar K, \Sigma_{b}^{*} \bar K) = \nonumber \\ 32.6\%,5.1\%,22.1\%,34.4\%,2.7\%,3.1\%
\end{eqnarray}
for the $\Xi_b^{\prime*}(2S)$ state. Clearly, the masses, total widths, and main decay modes of these two states are quite similar, and the secondary $\Xi^{\prime}_{b} \pi$ and $\Xi^{\prime*}_{b} \pi$ channels show significantly different behaviors. Thus, it is expected that the future experiments can search for these two states in the $\Xi_{b} \pi$ and $\Lambda_{b} \bar K$ final states and distinguish them from the $\Xi^{\prime}_{b} \pi$ and $\Xi^{\prime*}_{b} \pi$ decay modes.

Several candidates for the radial excited $6_F$ states of singly charmed baryons have been observed. Examples are the $\Sigma_c(2850)$, $\Xi_c(2965)/ \Xi_c(2970)$, and $\Omega_c(3090)$ structures; however, none of them has been confirmed as a $2S$ state either theoretically or experimentally. The situation is worse for the bottom sectors, since no signal for the radial excited $6_F$ state has been experimentally observed. More theoretical and experimental efforts are needed to discover and confirm these radially excited states.

\begin{table}[!htbp]
\begin{center}
\caption{\label{T2S} Theoretical predictions of the strong decays for the $\Xi^\prime_b(2S)$ and $\Xi^{\prime *}_b(2S)$ states in MeV.}
\renewcommand{\arraystretch}{1.5}
\begin{tabular*}{8.6cm}{@{\extracolsep{\fill}}*{3}{p{2.8cm}<{\centering}}}
\hline\hline
Mode &	$\Xi^{\prime}_{b}(2S)$	&	$\Xi^{\prime*}_b(2S)$\\\hline		
$\Xi_b \pi$	 &	11.06 	&	11.61 	\\
$\Xi^{\prime}_b \pi$	 &	6.49	&	1.80	\\
$\Xi^{\prime*}_b \pi$	 &	2.82	&	7.88 	\\
$\Lambda_b \bar K$	 &11.72	&	12.28	\\
$\Sigma_b \bar K$	 &	1.78	&	0.96	\\
$\Sigma^{*}_b \bar K$	 &	0.02	&	1.11 	\\
Total width	 &	33.89	&	35.64 \\
\hline\hline
\end{tabular*}
\end{center}
\end{table}

\subsection{$\Xi_b^\prime(1P)$ states}
From Table~\ref{bmass}, five $\lambda-$mode $\Xi_{b}^{\prime}(1P)$ states, which are denoted as $\Xi_{b0}^{\prime}(\frac{1}{2}^{-})$, $\Xi_{b1}^{\prime}(\frac{1}{2}^{-})$, $\Xi_{b1}^{\prime}(\frac{3}{2}^{-})$, $\Xi_{b2}^{\prime}(\frac{3}{2}^{-})$, and $\Xi_{b2}^{\prime}(\frac{5}{2}^{-})$, are allowed in the conventional quark model. The predicted masses of these states are in range of 6224 $\sim$ 6234 MeV, which show significantly small mass splitting. The strong decay behaviors of the five $\Xi_b^\prime(1P)$ states  calculated using these masses are listed in Table~\ref{B1P}. The pure $j=0$ state is too broad to be observed, while the two pure $j=1$ states can be searched in the $\Xi_b^\prime \pi$ and $\Xi_b^{\prime*} \pi$ final states. Further, the two $j=2$ states are relatively narrow and may be easily found in the $\Xi_b \pi$ and $\Lambda_b \bar K$ decay channels.

\begin{figure}[!htbp]
\includegraphics[scale=0.45]{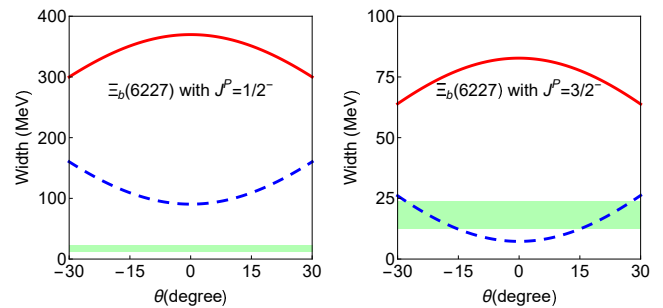}
\vspace{0.0cm} \caption{Total decay widths of $\Xi_b(6227)$ resonance under various assignments as functions of the mixing angle $\theta$ in the range $-30^\circ \sim 30^\circ$. The red solid lines are the $|1P~{1/2^-}\rangle_1$ and $|1P~{3/2^-}\rangle_1$ states, and the blue dashed curves correspond to the $|1P~{1/2^-}\rangle_2$ and $|1P~{3/2^-}\rangle_2$ states. The green bands indicate the measured total decay widths with errors.}
\label{xibp}
\end{figure}

Actually, the structure $\Xi_b(6227)$ is a good candidate for $\Xi_b^\prime(1P)$ states considering their mass and decay behaviors. Our calculations indicate that the assignments of $\Xi_b(6227)$ as pure $\Xi_b^\prime(1P)$ states are disfavored because the measured width of $\Xi_b(6227)$ is around 18 MeV~\cite{Aaij:2018yqz,Aaij:2020fxj}. However, the physical resonances may correspond to the mixtures of theoretical states with the same spin and parity quantum numbers. The mixing scheme of $P-$wave states can be expressed as
\begin{equation}
\left(\begin{array}{c}| 1P~{1/2^-}\rangle_1\cr | 1P~{1/2^-}\rangle_2
\end{array}\right)=\left(\begin{array}{cc} \cos\theta & \sin\theta \cr -\sin\theta &\cos\theta
\end{array}\right)
\left(\begin{array}{c} |1/2^-,j=0
\rangle \cr |1/2^-,j=1\rangle
\end{array}\right),
\end{equation}
\begin{equation}
\left(\begin{array}{c}| 1P~{3/2^-}\rangle_1\cr | 1P~{3/2^-}\rangle_2
\end{array}\right)=\left(\begin{array}{cc} \cos\theta & \sin\theta \cr -\sin\theta &\cos\theta
\end{array}\right)
\left(\begin{array}{c} |3/2^-,j=1
\rangle \cr |3/2^-,j=2\rangle
\end{array}\right).
\end{equation}
Under the heavy quark limit, the mixing angle should be zero. Given the finite mass of the bottom quark, the heavy quark symmetry should be approximately preserved, and the mixing angle may show a small divergence between the physical structures and theoretical states in $j-j$ couplings. The total decay widths of various assignments for $\Xi_b(6227)$ versus the mixing angle $\theta$ in the range of $-30^{\circ} \sim 30^{\circ}$ are plotted in Fig.~\ref{xibp}. When the mixing angle $\theta$ lies in $12^{\circ} \sim 29^{\circ}$ (or $-12^{\circ} \sim -29^{\circ}$), the strong decay width is consistent with the experimental data. This mixing angle is close to that obtained in our previous work for the singly bottom baryon $\Sigma_b(6097)$~\cite{Liang:2019aag}, which is the non-strange partner of the $\Xi_b(6227)$ resonance. The sign of the mixing angle cannot be determined by the strong decay behaviors, and more theoretical efforts are needed to clarify this problem in the future. With the mixing angle $\theta$=$22^{\circ}$, the branching ratios for the $\Xi_b(6227)$ resonance are predicted as
\begin{equation}
Br(\Xi_{b} \pi, \Xi^{\prime}_{b} \pi, \Xi^{\prime*}_{b} \pi, \Lambda_{b} \bar K) = 21.8\%,2.3\%,65.1\%,10.8\%.
\end{equation}

Our present assignment for the $\Xi_b(6227)$ is consistent with the results of previous works within the potential model~\cite{Chen:2018orb}, chiral quark model~\cite{Wang:2018fjm}, Regge approach~\cite{Jia:2019bkr}, and QCD sum rule~\cite{Cui:2019dzj,Azizi:2020azq}, and the theorists too seemed to have reached a consensus on this state. However, interestingly, all the $\Xi_b^\prime(1P)$ states can decay into the $\Xi_b \pi$ and $\Lambda_b \bar K$ channels under the mixing scheme. Except for the broad $|1P~{1/2^-}\rangle_1$ state, four other states with similar masses, are most likely to be observed in the $\Xi_b \pi$ and $\Lambda_b \bar K$ invariant masses. This contradicts the present experimental observation where only one structure $\Xi_b(6227)$ was seen. Hence, it is possible that the structure $\Xi_b(6227)$ corresponds to more than one resonance. Future experiments may reanalyze the structure $\Xi_b(6227)$ to clarify this puzzle.

\begin{table}[!htbp]
\begin{center}
\caption{\label{B1P}Theoretical predictions of the strong decays for the $\Xi^{\prime}_{b}(1P)$ states in MeV.}
\renewcommand{\arraystretch}{1.5}
\footnotesize
\begin{tabular*}{8.6cm}{@{\extracolsep{\fill}}*{6}{p{1.3cm}<{\centering}}}
\hline\hline
 Mode&	$\Xi^{\prime}_{b0}(\frac{1}{2}^-)$&$\Xi^{\prime}_{b1}(\frac{1}{2}^-)$&	$\Xi^{\prime}_{b1}(\frac{3}{2}^-)$	&	$\Xi^{\prime}_{b2}(\frac{3}{2}^-)$	&	$\Xi^{\prime}_{b2}(\frac{5}{2}^-)$	 \\\hline
$\Xi_b \pi$	 &	197.69	&	$-$&	$-$ &4.38&		 4.50\\
$\Xi^{\prime}_b \pi$	 &	$-$ &90.30	 &0.19 &0.27& 0.13\\
$\Xi^{\prime *}_b \pi$	 &	$-$ &0.21 &	85.99 &0.18 &0.29	\\
$\Lambda_{b} \bar K$	 &	173.06&	$-$	&	$-$ &2.10 &2.20	\\
Total width	 &	370.75&90.51&86.18 &6.93&7.12 \\
\hline\hline
\end{tabular*}
\end{center}
\end{table}

\subsection{$\Xi_b^\prime(1D)$ states}

From Table~\ref{bmass}, the masses of six $\Xi^{\prime}_b(1D)$ states are predicted to be in the range of $6414\sim 6459$ MeV. The strong decay widths for these $D-$wave states are listed in Table~\ref{T1D}. The total widths of $\Xi^{\prime}_{b1}(\frac{1}{2}^+)$, $\Xi^{\prime}_{b1}(\frac{3}{2}^+)$, $\Xi^{\prime}_{b2}(\frac{3}{2}^+)$, $\Xi^{\prime}_{b2}(\frac{5}{2}^+)$, $\Xi^{\prime}_{b3}(\frac{5}{2}^+)$, and $\Xi^{\prime}_{b3}(\frac{7}{2}^+)$ states are about 39, 40, 22, 19, 2, and 2 MeV, respectively. For the two $j=1$ states, the dominating decay modes are $\Xi_b \pi$ and $\Lambda_b \bar K$, while other decay channels are relatively small. For the two $j=2$ states, the $\Xi_b \pi$ and $\Lambda_b \bar K$ final states are not allowed because of the spectator assumption in the $^3P_0$ model. Although the masses and total decay widths of these two states are similar, the main decay modes are significantly different; this is helpful to differentiate the two states. For the two $j=3$ states, the calculated decay widths are about several MeV, and the main decay modes are $\Xi_b \pi$ and $\Lambda_b \bar K$. Moreover, owing to the higher initial masses, the $\Xi_b^\prime(1D)$ states may also decay into the $\Lambda \bar B$ and $\Lambda \bar B^*$ final states. These light baryon–heavy meson decay modes may be important for the higher excited $\lambda-$mode states~\cite{Lu:2019rtg} but can be neglected for the low-lying states owing to limited phase space. We expect that the present predictions are helpful for future experiments to search for these $\Xi^{\prime}_b(1D)$ states.

\begin{table}[!htbp]
\begin{center}
\caption{ \label{T1D}Theoretical predictions of the strong decays for the $\Xi^{\prime}_b(1D)$ states in MeV.}
\renewcommand{\arraystretch}{1.5}
\footnotesize
\begin{tabular*}{8.6cm}{@{\extracolsep{\fill}}p{1.3cm}<{\centering}p{1cm}<{\centering}p{1cm}<{\centering}p{1cm}<{\centering}p{1cm}<{\centering}p{1cm}<{\centering}p{1cm}<{\centering}}
\hline\hline
Mode	&	$\Xi^{\prime}_{b1}(\frac{1}{2}^+)$	&	$\Xi^{\prime}_{b1}(\frac{3}{2}^+)$	&	$\Xi^{\prime}_{b2}(\frac{3}{2}^+)$	&	$\Xi^{\prime}_{b2}(\frac{5}{2}^+)$	&	$\Xi^{\prime}_{b3}(\frac{5}{2}^+)$	&	 $\Xi^{\prime}_{b3}(\frac{7}{2}^+)$	\\\hline
$\Xi_b \pi$	 &13.85 &14.10 &$-$ &$-$ &0.94&0.88	\\
$\Xi^\prime_b \pi$	 &3.04 &0.80 &6.34 &0.10	&0.09&0.05\\
$\Xi^{\prime*}_b \pi$ &1.41 &3.72&1.30 &7.09&0.09&0.11\\
$\Lambda_{b} \bar K$	 &12.32	 &12.30 &$-$&$-$&1.22&1.13\\
$\Sigma_b \bar K$	 &6.19	 &1.71 &11.92&0.07&0.05&0.02\\
$\Sigma^{*}_b \bar K$	 &2.54 &7.19 &1.94&11.49&0.03&0.03\\
Total width &39.35 &39.82 &21.50 &18.75&2.42&2.23\\	
\hline\hline
\end{tabular*}
\end{center}
\end{table}

\section{summary}{\label{summary}}

In this work, we investigated the strong decays for the $\lambda-$mode low-lying $\Xi_{b}$ and $\Xi_{b}^\prime$ baryons within the $^3P_0$ model. Considering the masses and strong decay behaviors, we can assign the newly observed $\Xi_{b}(6100)$ resonance as the $\Xi_b(1P)$ state with $J^P=3/2^-$ and interpret the $\Xi_{b}(6227)$ structure as the $J^P=3/2^-$ $\Xi_b^\prime(1P)$ state with a proper mixing angle. According to our predictions, all of the low-lying $\bar 3_F$ states $\Xi_b(2S)$, $\Xi_b(1P)$, and $\Xi_b(1D)$ have narrow decay widths of less than 20 MeV, and they are most likely to be observed in the $\Xi_b^\prime \pi$ and $\Xi_b^{\prime *} \pi$ invariant masses. Further, the total decay widths for most of the $6_F$ multiplets are relatively small, and these multiplets may decay into the $\Xi_b \pi$, $\Xi_b^\prime \pi$, $\Xi_b^{\prime *} \pi$, and $\Lambda_b \bar K$ final states.

Although the $\Xi_b$ and $\Xi_b^\prime$ systems are more complicated, they show similar mass patterns and decay behaviors compared with other singly heavy baryons. The light flavor SU(3) symmetry and heavy quark symmetry may be preserved well in these systems and provide valuable clues to establish the low-lying spectra. The purpose of this work is not to give accurate predictions of the strong decay behaviors for the excited $\Xi_b$ and $\Xi_b^\prime$ states but to identify the possible narrow states and typical decay modes. In this regard, the present results are expected to be helpful for future experimental searches. Considering the uncertainties in the adopted masses and the $^3P_0$ model, one cannot expect the estimated decay widths to be accurate. Empirically, these predictions contain uncertainties of about $30\%$ and can be regarded as semiquantitative estimations~\citep{Lu:2019rtg}. More theoretical and experimental efforts for the $\Xi_b$ and $\Xi_b^\prime$ states are needed to establish their low-lying spectra.

\acknowledgments

We would like to thank Xian-Hui Zhong for valuable discussions. This work is supported by the National Natural Science Foundation of China under Grants No. 11705056,
No. 11975245, No. U1832173, and No. 11475192. This work is also supported by the Sino-German CRC 110 “Symmetries and the Emergence of Structure in QCD” project by NSFC
under the Grant No. 12070131001, the Key Research Program of Frontier Sciences, CAS, under the Grant No. Y7292610K1, and the National Key Research and Development Program of
China under Contracts No. 2020YFA0406300.

\end{document}